\documentclass[a4paper,12pt]{article}
\usepackage{graphicx}
\usepackage{authblk}
\usepackage{amsmath}
\usepackage[margin=1in]{geometry}
\title{Computational spectral-domain single-pixel imaging}
\date{}

\author[1]{\small Piotr Ryczkowski}
\author[1]{\small Caroline Amiot}
\author[2]{\small John M. Dudley}
\author[1]{\small Go\"ery Genty}
\affil[1]{\footnotesize Laboratory of Photonics, Tampere University, FI-33101 Tampere, Finland}
\affil[2]{\footnotesize Institut FEMTO-ST, Universit\'{e} Bourgogne Franche-Comt\'{e} CNRS UMR 6174, 25000 Besan\c{c}on, France}
\begin{document}

\maketitle

\begin{abstract}
We demonstrate single-pixel imaging in the spectral domain by encoding Fourier probe patterns onto the spectrum of a superluminescent laser diode using a programmable optical filter. As a proof-of-concept, we measure the wavelength-dependent transmission of a Michelson interferometer and a wavelength-division multiplexer. Our results open new perspectives for remote broadband measurements with possible applications in industrial, biological or security applications. 
\end{abstract}

\section{Introduction}
Ghost imaging is an indirect measurement technique that uses the correlation between the intensity profile of a spatially-resolved light beam and the spatially-integrated intensity of the same beam transmitted through (or reflected from) an object to reconstruct an image of that object \cite{Sun2019,Edgar2019}. When first developed, the illumination patterns were based on random noise, which requires the illuminating beam intensity profile must be measured in a separate reference arm \cite{Erkmen2010}. The image is then obtained from the correlation of the measured reference beam profiles and signals provided by a single pixel (integrating) detector placed after the object. Extensively studied to reconstruct spatial images of physical objects, the technique has recently been extended to other domains including temporal \cite{Ryczkowski2016,O-Oka2017,Devaux2016} and spectral domains \cite{amiot2018supercontinuum,amiot2019ghost,Janassek2018b}. 

The need for distinct reference measurements of the illuminating patterns can be eliminated using a computational approach also commonly referred to as single-pixel imaging  \cite{Erkmen2010}. In this case, a set of specially designed intensity masks (that do not need to be measured as they are computationally stored) are used for illumination, and it is then only necessary to use just one single-pixel detector to measure the integrated intensity after light interaction with the object. The image can then be reconstructed by solving a simple inversion problem. Using an appropriate set of (mathematically) orthogonal illuminating patterns, the single-pixel imaging approach is significantly faster and yields enhanced signal-to-noise ratio as compared to conventional random illumination ghost imaging. The technique is also simpler as no reference measurement is needed. Moreover, if the measured object is sparse, one can further use compressed sensing schemes to reduce even further the number of distinct measurements \cite{Romberg2008,Wenwen2019}. Single-pixel imaging has been demonstrated both in the spatial and in the temporal domain \cite{Erkmen2010,Devaux2016,Zhang2015a}.

Here, we propose and demonstrate single-pixel imaging in the spectral domain using a continuous broadband light source whose spectrum is modulated with harmonic series of sine and cosine patterns \cite{Wenwen2019,Salvador-Balaguer2016,Zhang2017}. As previously demonstrated \cite{amiot2018supercontinuum,amiot2019ghost}, spectral ghost imaging allows the acquisition of data such as absorption spectra or images. The advantage of this approach is that one obtains directly the Fourier Transform coefficients of the object's spectral response which can then be retrieved by simple inversion. This leads to significant speed improvement compared to with using a light source with random spectral fluctuations necessitating the use of a reference arm and subsequent correlation \cite{amiot2018supercontinuum,amiot2019ghost}. The technique is especially adapted to remote sensing when spectrally-resolved measurements are not possible e.g. in the presence of strong scattering or low signal levels, and it also can be extended to multi-dimensional hyperspectral measurements \cite{Schultz2001,Johnson2007,Lorente2012,Elmasry2012,Sorg2005,Hagen2013,ElMasry2007,Kawakami2011,Kaariainen2019}. 

We begin by describing the general principle of spectral-domain single-pixel imaging. Consider an object with a specific spectral response (transmission or reflection) $T(\omega_0+\Omega)$, where $\Omega$ is a relative optical frequency, spanning $\Delta\omega$ around a central frequency $\omega_0$. The spectral response of the object can be decomposed onto a basis of (truncated) harmonic series of $N+1$ sine and cosine functions such that:
\begin{equation}
T(\omega_0+\Omega)=  \sum_{n=0}^{N} a_{n}\cos\left({\frac{2\pi n \Omega}{\Delta\omega}}\right) + \sum_{n=0}^{N} b_{n}\sin\left({\frac{2\pi n \Omega}{\Delta\omega}}\right),
\label{eq:1}
\end{equation}
where $a_{n}$ and $b_{n}$ represent the $n^{th}$ cosine and sine Fourier coefficients, respectively, defined as
\begin{align}
a_{n} = \frac{2}{\Delta\omega}\int_{-\frac{\Delta\omega}{2}}^{+\frac{\Delta\omega}{2}}T(\omega_0+\Omega)\cos\left({\frac{2\pi n \Omega}{\Delta\omega}}\right) d\Omega\\
b_{n} = \frac{2}{\Delta\omega}\int_{-\frac{\Delta\omega}{2}}^{+\frac{\Delta\omega}{2}}T(\omega_0+\Omega)\sin\left({\frac{2\pi n \Omega}{\Delta\omega}}\right) d\Omega.
\end{align}
Therefore, by illuminating the object with a light source whose spectral intensity is modulated by sinusoidal patterns of different angular frequencies and measuring with a single-pixel detector the (spectrally) integrated intensity after transmission (or reflection) through the object, one can obtain the Fourier coefficients and thereby retrieve the object's spectral response from Eq. (\ref{eq:1}). In practice, the modulation is encoded onto the spectral intensity of the light source, which means that the sinusoidal modulation (sine or cosine) has a DC component equal to source's mean spectral intensity $I_{0}$. In order to eliminate the DC component, the object can be illuminated with complementary modulation patterns with reversed phase \cite{Zhang2015a,Yu2015}:
\begin{align}
I_n^{\pm c}(\omega_0+\Omega)=I_{0}\left[1 \pm \cos\left({\frac{2\pi n \Omega}{\Delta\omega}}\right)\right] \\ 
I_n^{\pm s}(\omega_0+\Omega)=I_{0}\left[1 \pm \sin\left({\frac{2\pi n \Omega}{\Delta\omega}}\right)\right],
\end{align}
and the spectral response is then retrieved from:
\begin{equation}
T(\omega_0+\Omega) =  \frac{1}{2I_{0}}
	\sum_{n=0}^{N} \left(a_{n}^{+}-a_{n}^{-}\right)\cos\left({\frac{2\pi n \Omega}{\Delta\omega}}\right) +
\frac{1}{2I_{0}}
 	\sum_{n=0}^{N} \left(b_{n}^{+}-b_{n}^{-}\right)\\sin\left({\frac{2\pi n \Omega}{\Delta\omega}}\right),
\label{eq:2}
\end{equation}
where
\begin{align}
a_{n}^{\pm} = \frac{2}{\Delta\omega}\int_{-\frac{\Delta\omega}{2}}^{+\frac{\Delta\omega}{2}}T(\omega_0+\Omega)I_n^{\pm c}(\omega_0+\Omega) d\Omega\\
b_{n}^{\pm} = \frac{2}{\Delta\omega}\int_{-\frac{\Delta\omega}{2}}^{+\frac{\Delta\omega}{2}}T(\omega_0+\Omega)I_n^{\pm s}(\omega_0+\Omega) d\Omega.
\end{align}
A total of 4N+2 patterns are then required to reconstruct the spectral transmission of the object (sine patterns for $n=0$ can be omitted), and the fact that one uses a truncated series limits the measurement spectral resolution to $\Delta\omega/\rm N$. Note that this constitutes a lower limit for the resolution and in practice the resolution can be further degraded due to the due to the limited bandwidth of the programmable filter resulting in a loss of contrast of the probing patterns contrast at higher modulation frequencies. 
\section{Experimental setup}
\begin{figure}[h!]
	\begin{center}
	\includegraphics[width=14 cm ]{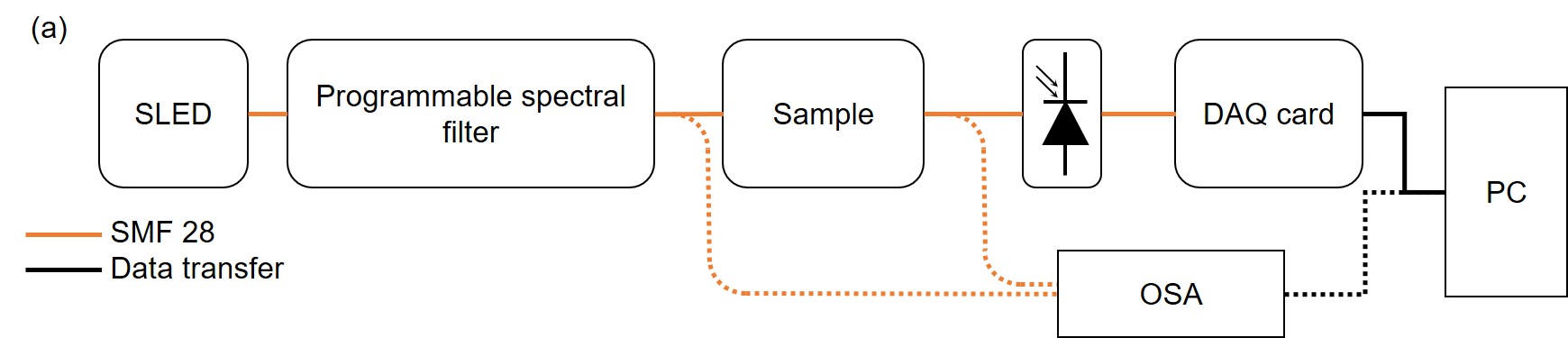}
	\includegraphics[width=5 cm ]{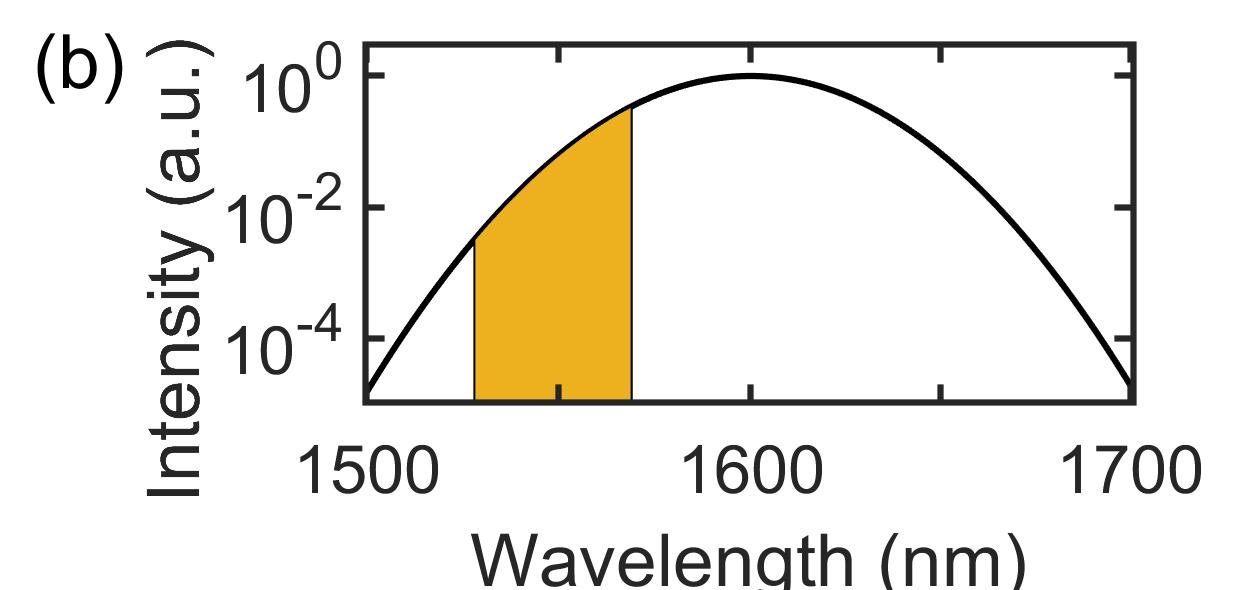}
	\includegraphics[width=10 cm]{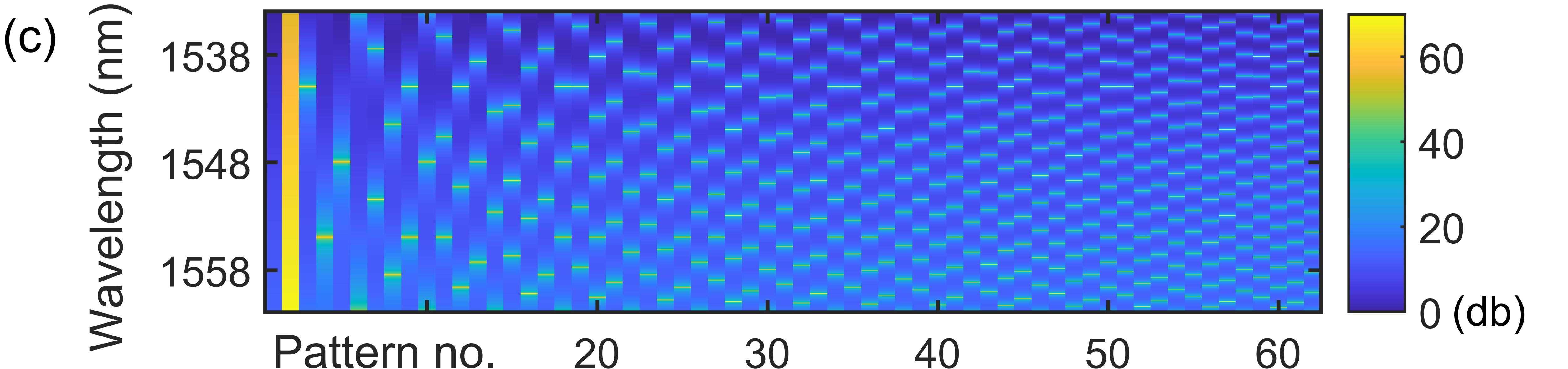}
	\end{center}
	\vspace*{-6mm}
	\caption{(a) Experimental setup. (b) SLED full spectrum with the wavelength band of the programmable spectral filter highlighted in orange. (c) Programmed spectral filter attenuation to generate the complementary Fourier pairs, here limited to 33 pairs patterns (N=16) for clarity. The patterns are plotted in the following order: constant intensity ($I_0^{+c}$), zero intensity ($I_0^{-c}$), followed by complementary pairs of sine and cosine with increasing modulation frequency.}
	\label{fig:setup}
\end{figure}
Our experimental setup is shown in Fig.~\ref{fig:setup}a. Light from a fiber-coupled superluminescent diode (SLED) (Exalos ESL1620-2111) is directed through a programmable spectral filter (Finisar Waveshaper 4000s) which sequentially modulates the spectral intensity according to the complementary patterns described above (Fig. \ref{fig:setup}c). The output of the programmable filter is collimated (or fiber-coupled) to illuminate the sample under test. Light after the sample is recorded with a single-pixel large area detector (Thorlabs PDA50B-EC) with no spectral resolution. The electronic signal corresponding to each sinusoidal pattern is digitized with a DAQ card (NI USB-6212) and stored in a computer. The sequential feeding of modulated spectral patterns and data acquisition is controlled with a Labview program. The spectral response is then reconstructed by post-processing using Eq. (6). Note that it is straightforward to compensate for wavelength-dependence of the source by by pre-normalizing the probing sinusoidal patterns to the SLED unmodulated spectrum (Fig. \ref{fig:setup}b). In order to validate the method, we also perform an independent reference measurement of the sample spectral response using an optical spectrum analyser (OSA,  Ando AQ-6315B). 

\begin{figure}[h!]
	\begin{center}
	\includegraphics[width=7 cm ]{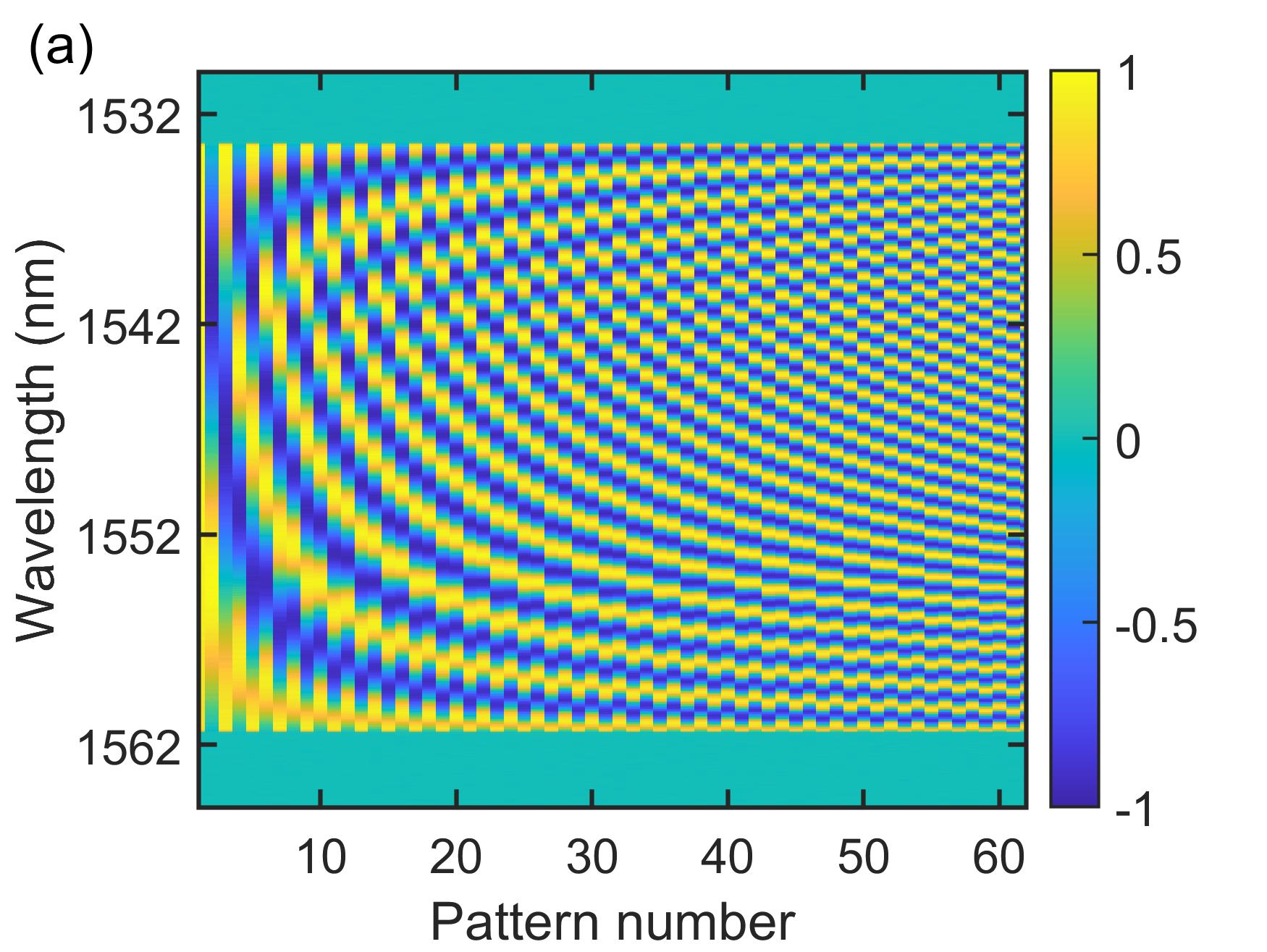}
	\includegraphics[width=7 cm ]{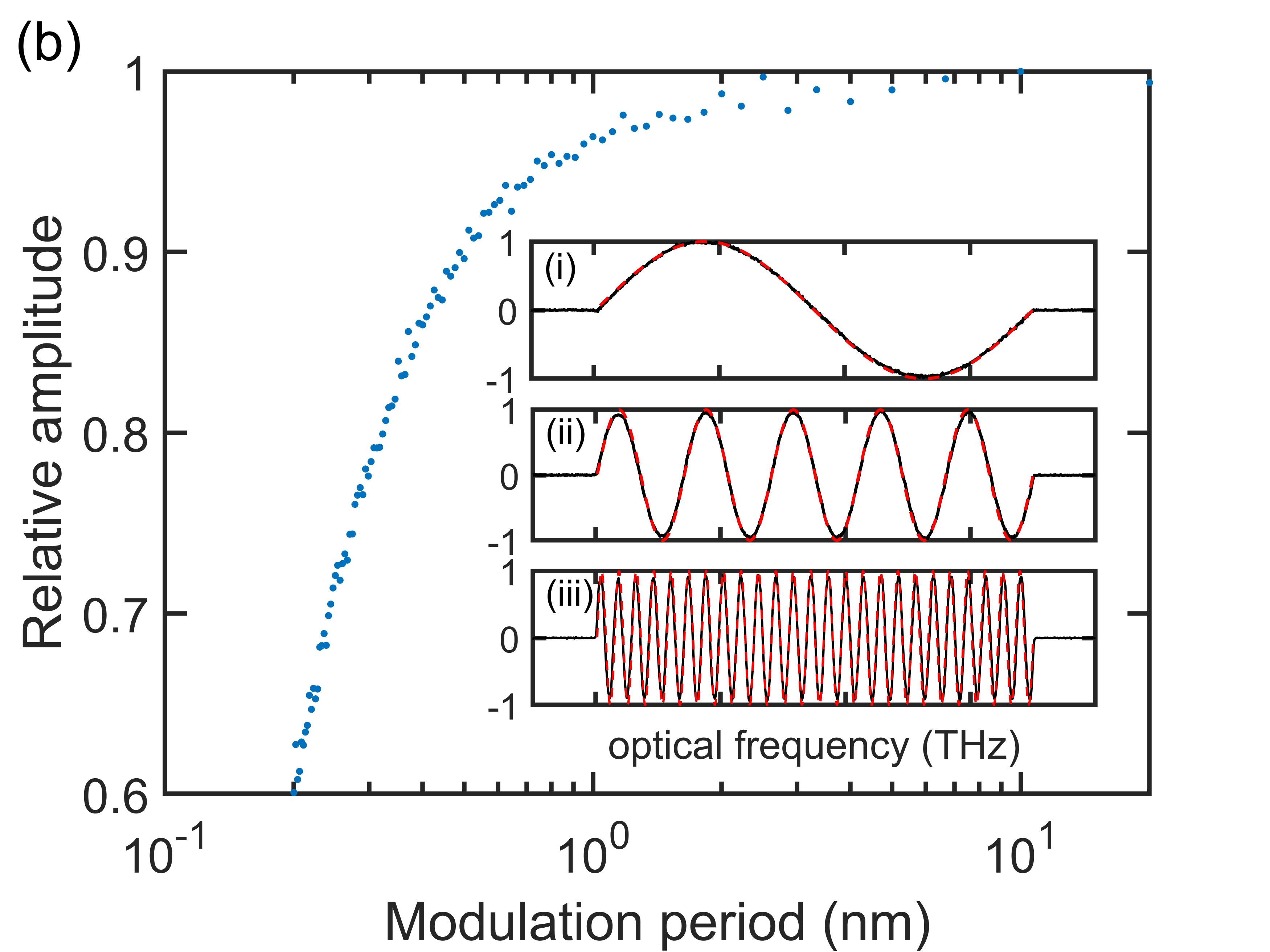}
\end{center}
\vspace*{-6mm}
	\caption{(a) First 61 pure sine and cosine patterns derived from the complementary pairs measured individually with the OSA. The fist pattern represents the unmodulated SLED spectrum (constant pattern) followed by pairs of sine and cosine patterns of increasing modulation frequency (decreasing modulation period). (b) Modulation amplitude of the 100 pure sine patterns as a function of modulation period, normalized to the highest amplitude value of the slowest modulated pattern. For comparison, the inset shows the measured patterns for three specific modulation frequencies 3.5~THz ($\sim$28~nm), 0.7~THz ($\sim$5.6~nm), and 0.14~THz ($\sim$1.1~nm) as solid black lines  together with ideal sine functions as dashed red lines.}
	\label{fig:reference_patterns}
\end{figure}

The experimentally measured pre-programmed patterns with the OSA are compared against ideal sinusoidal modulations in Fig. \ref{fig:reference_patterns}. Although in the measurements reported below we used a total of 402 patterns (N=~100), here for clarity we only plot the first 61 patterns (including the unmodulated SLED spectrum plus the first 30 sine and cosine modulations). One can see how the period of all the programmed spectral modulations match very well with ideal spectral sinusoidal modulations. We also do observe a loss in modulation contrast as the modulation frequency is increased which is caused by the limited operation bandwidth of the programmable filter and this is the main limitation for the measurement resolution. 

\section{Results}
\subsection{Object with periodic spectral transmission}
\begin{figure}[hbt!]
	\begin{center}
		\includegraphics[width=7 cm ] {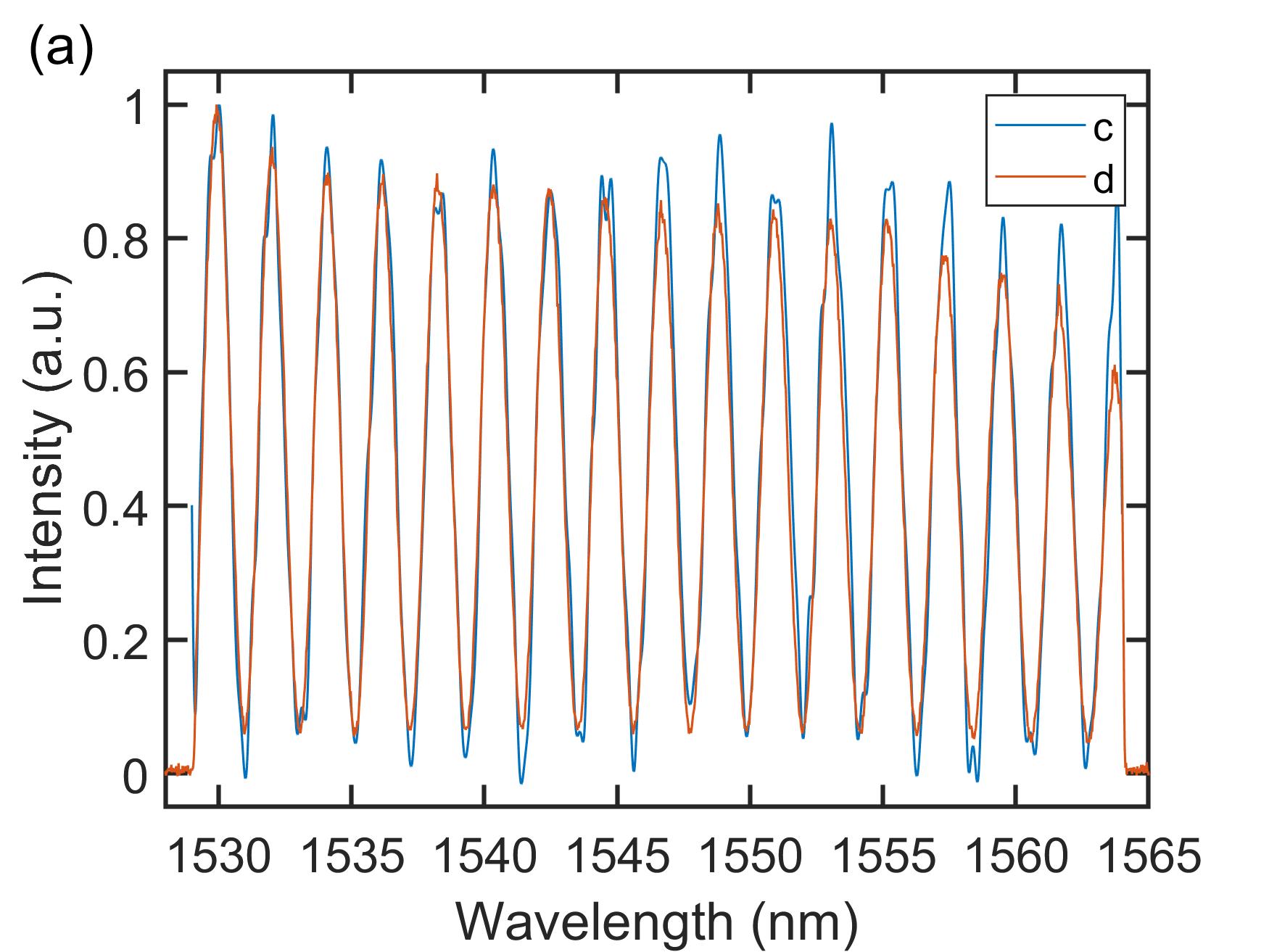}
		\includegraphics[width=7 cm ] {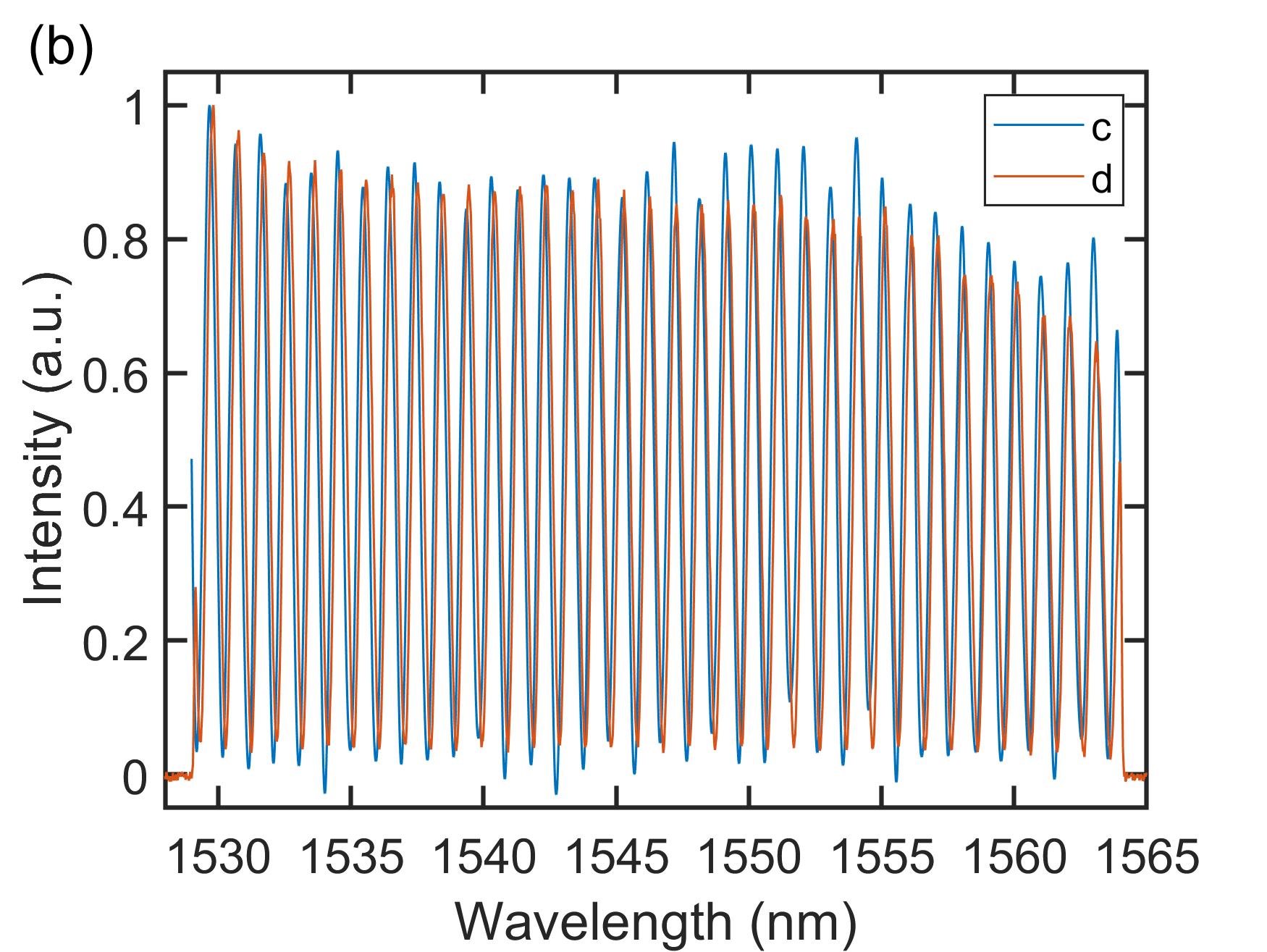}
		\includegraphics[width=7 cm ] {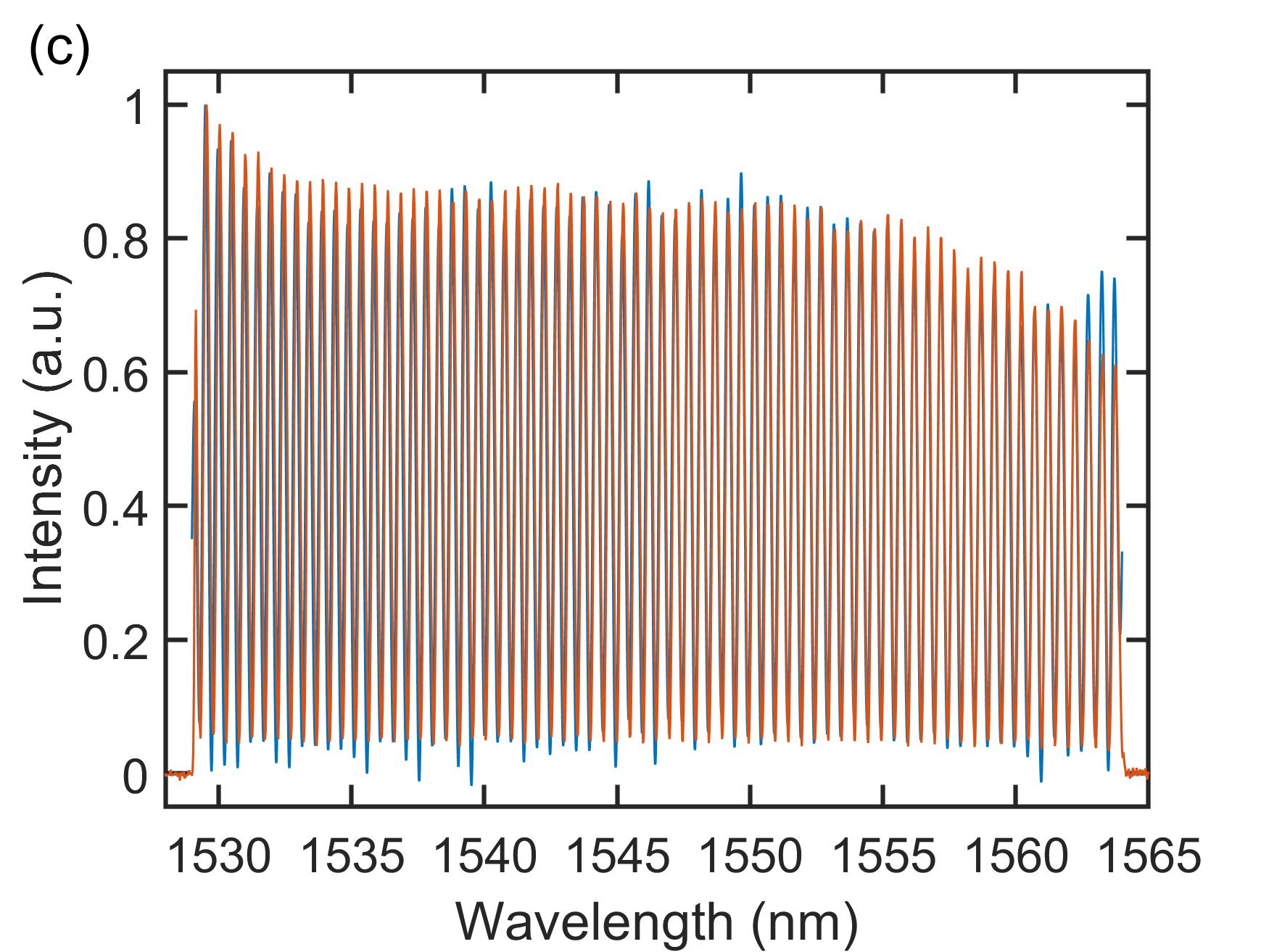}
		\includegraphics[width=7 cm ] {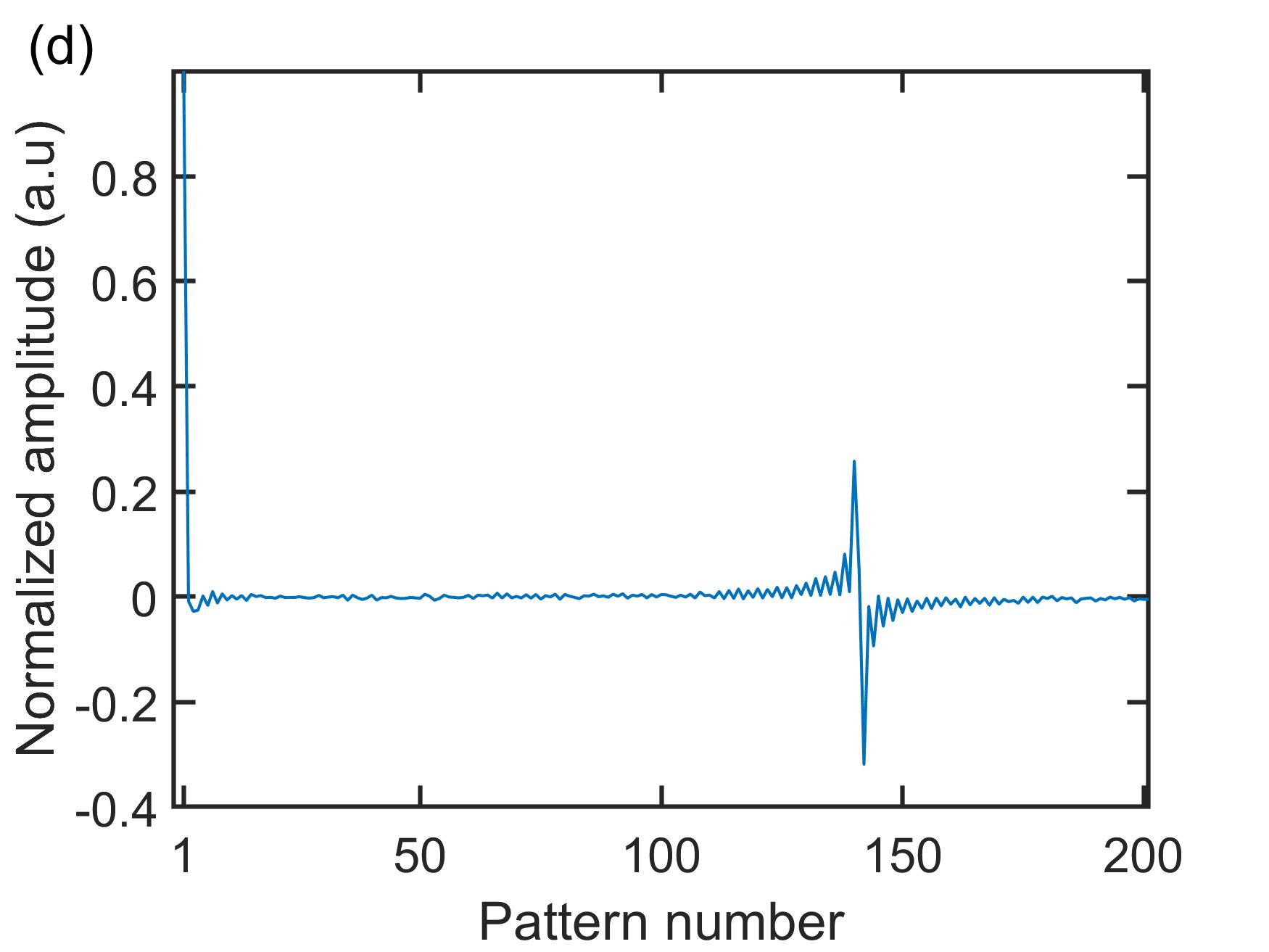}
	\end{center}
	\vspace*{-6mm}
	\caption{Measured transmission spectra of an unequal path Michelson interferometer with 4~ps (a), 8~ps (b) and 16~mm (c) temporal delay (optical path differences of 1.2~mm, 2.4~mm and 4.8~mm, respectively) corresponding to a spectral modulation period of 2~nm, 1~nm, and 0.5~nm, respectively. The blue and red solid lines represent the single-pixel imaging measurements and direct OSA measurement, respectively. (d) shows the signal measured by the integrating single-pixel detector as a function of the modulation pattern frequency for the case of 16~ps delay.}
	\label{fig:FabryPerot_examples}
\end{figure}

We first probed the spectral response of an unequal path Michelson interferometer known to be a purely sinusoidal function of frequency with modulation period inversely proportional to the temporal delay between the two interferometer arms. In this case, one expects that only the probing sinusoidal patterns (sine and cosine) with modulation period corresponding to the temporal delay between the two arms will be transmitted at the interferometer output and result in a high intensity signal on the single-pixel detector. In order to verify this, we conducted a series of measurements where we probed the spectral transmission of the Michelson interferometer with complementary Fourier patterns for different optical path differences between the two arms. In this set of measurements, we have used $N=100$ and 36~nm bandwidth (1528~nm to 1564~nm), corresponding to an effective spectral resolution of 0.36 nm. The DAQ card operated at 1~MHz sampling rate with 1~s averaging for each modulated pattern. Note that the measurement speed is further limited by the switching time in the programmable filter (500~ms between consecutive patterns) over which data collection is held. The results are shown in Fig. \ref{fig:FabryPerot_examples} for temporal delay of 4~ps (a), 8~ps (b) and 16~ps (c). The amplitude of the signal as measured by the single-pixel detector, in case of 16~ps delay, for all patterns is also plotted in the figure (see Fig. \ref{fig:FabryPerot_examples}d). One can see that for all the different optical path differences, the modulation period and phase are correctly retrieved and that, as expected, one only observes high signal at the single-pixel detector for a modulation frequency corresponding to the inverse of the optical path difference between the two arms of the interferometer.  We also note the presence of additional components caused by the non-uniform transmission of the interferometer but these can be also be seen in the reference OSA measurement.

\subsection{Object with continuous spectral transmission}
\begin{figure}[]
	\begin{center}
		\includegraphics[width=7 cm ]{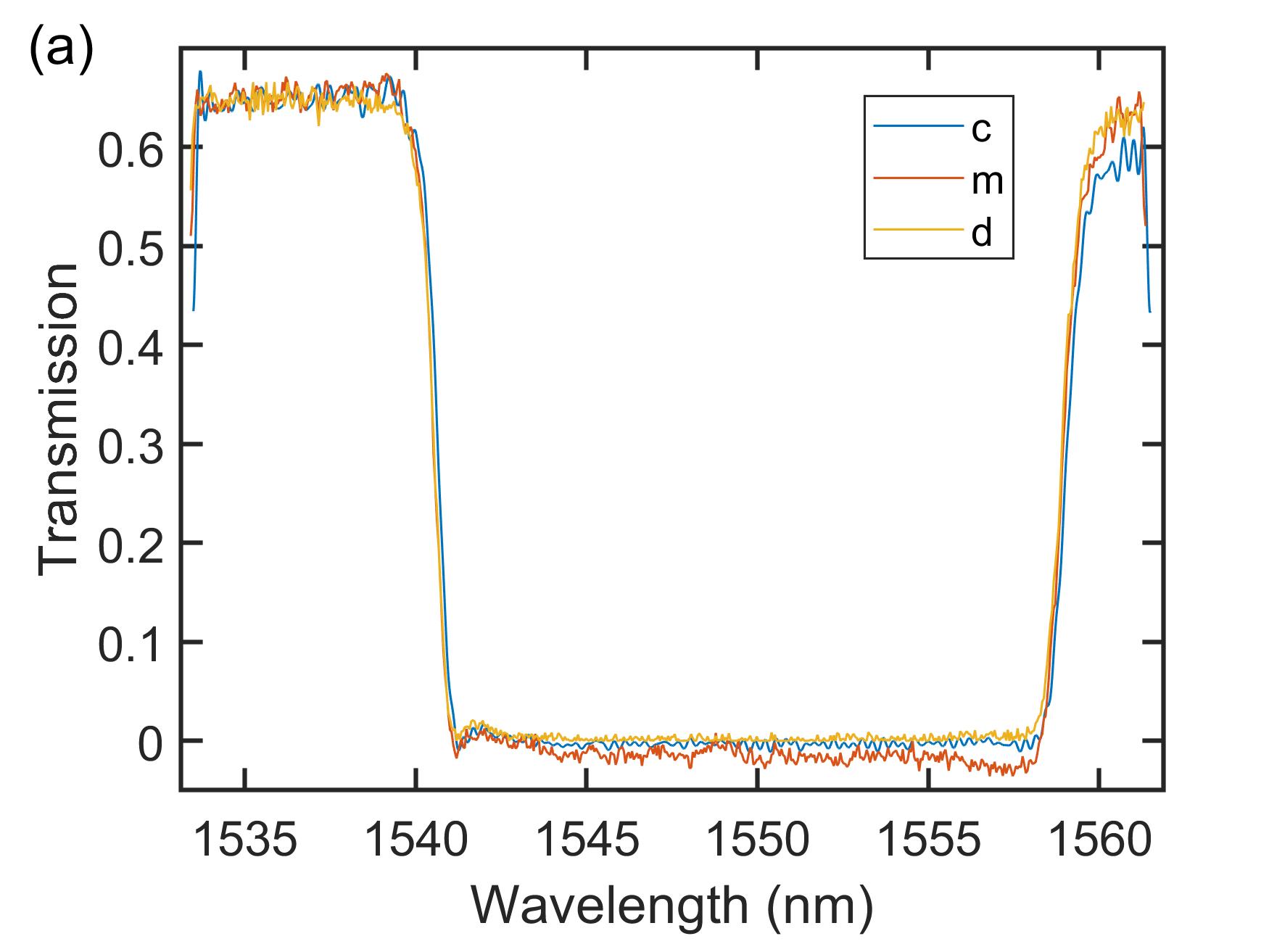}
		\includegraphics[width=7 cm ]{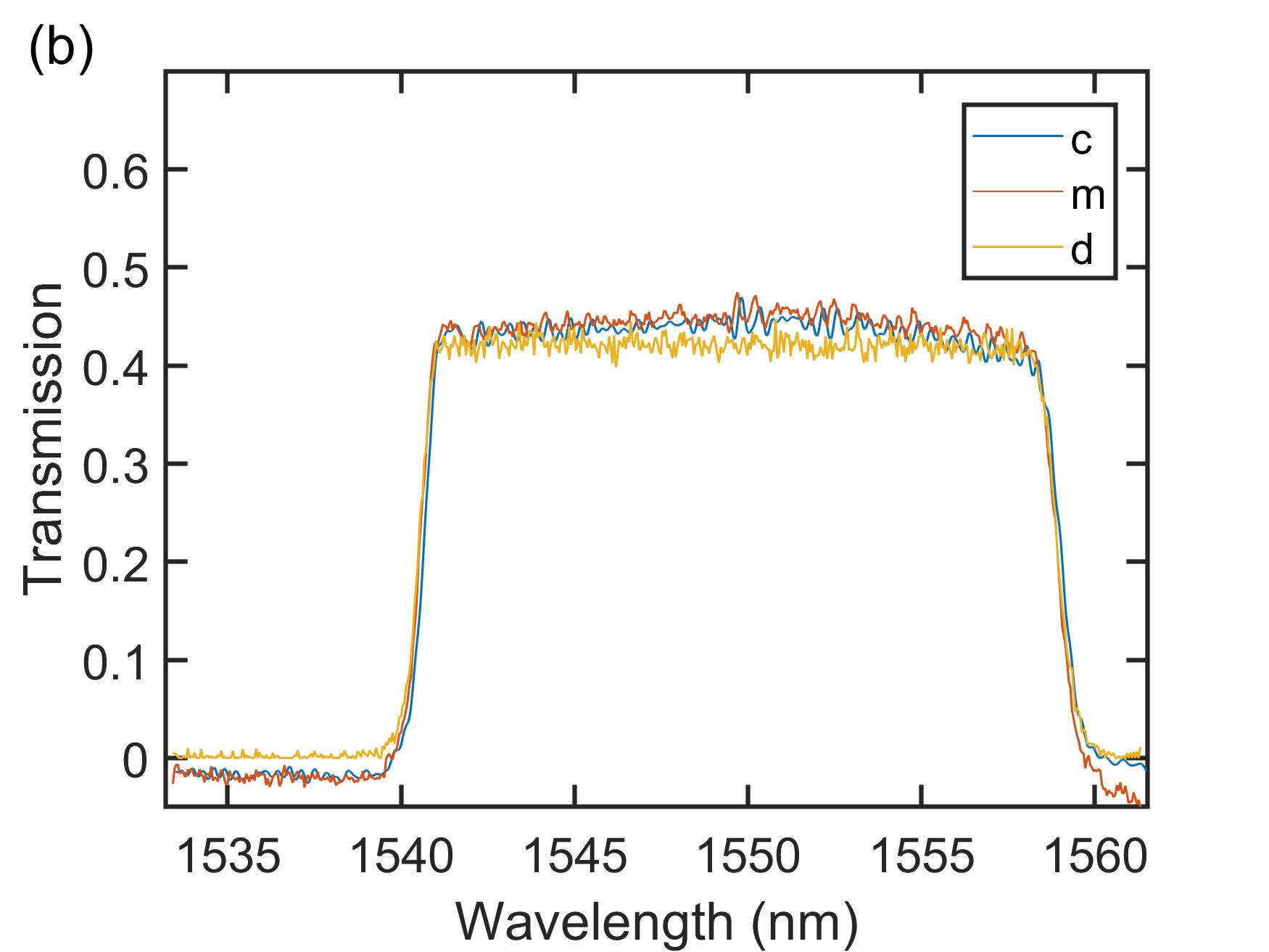}
	\end{center}
	\vspace*{-6mm}
	\caption{Transmission of a wavelength-division multiplexer measured in rejection (a) and transmission (b) configurations. The solid lines marked as "c","m" and "d" correspond to the spectral transmission obtained from single-pixel measurements using ideal complementary patterns, single-pixel measurements using the measured complementary patterns at the waveshaper output, and direct measurement using the OSA, respectively.}
	\label{fig:band_and_gap_example}
\end{figure}

We next performed single-pixel imaging measurements of the output ports of a wavelength-division multiplexer which exhibit sharp spectral features. Note that here we limited the SLED spectrum to 28~nm bandwidth from 1534~nm to 1562~nm while keeping $N=100$, corresponding to an effective spectral resolution of 0.28 nm. The results are shown in Fig.\ref{fig:band_and_gap_example} using the pre-measured complementary Fourier patterns along with a reference measurement performed with the OSA for comparison. For completeness, we also show in the figure the single-pixel imaging results obtained using ideal sine and cosine functions (i.e. not pre-measured) to reconstruct the spectral transmission of the two ports. We can see very good agreement between both single pixel technique approaches (i.e. using ideal mathematical functions or pre-measured patterns) with the OSA reference measurement in terms of amplitude and bandwidth as well as at the band edges where the transmission slope is steepest.  This means that in principle one does not even need to pre-measure and store the Fourier patterns in a computer and the reconstruction can be simply performed using theoretical complementary functions. We do note a slight increase in the noise amplitude of the single-pixel measurements as compared to that of the OSA which we attribute to the accumulated error during the spectral response retrieval calculation. The fact that one uses as truncated Fourier series expansion can also artificially smooth the retrieved spectral response and this can be seen when comparing the residual modulation on top of the stop/pass band as measured by the OSA and that retrieved from the single-pixel imaging measurement. 

\section{Discussion and conclusions}
We have demonstrated single-pixel imaging in the spectral domain using complementary Fourier-pair probing patterns. The technique allows for measurement of the Fourier coefficients of the object's spectral response with fast reconstruction. It also intrinsically eliminates the dark current of the detection system and potential parasitic influence of the ambient light irradiating the object during the measurements. The resolution of the technique is limited by the resolution of the spectral filter and/or the number of probing patterns. In contrast with direct measurement techniques that use wavelength-sensitive detection, here the weight of the measurement is placed on the illumination pattern relaxing significantly the constraint on the detector sensitivity. This can be particularly interesting from the perspective of remote hyperspectral measurement of physical objects where once could, in principle, acquire different images corresponding to different spectral illumination patterns using simply an unfiltered camera. The technique can be extended to any wavelength region and can be implemented with relatively inexpensive components. Although here we used complementary Fourier pair spectral illumination patterns, one can in principle also use different types of patterns using e.g. Hadamard functions and which could be more suitable for specific applications. Using digital mirror arrays in place of liquid crystal-based programmable filter would greatly increase the detection speed. Sensitivity of the measurements may also be increased by adding lock-in detection to the current setup. Our result could open up new applications for remote spectral measurements in industrial, biological or security applications, where one illuminates a specific target with pre-programmed spectral patterns and detects the retro-reflected signal with an integrating detector without any spectral resolution.

\section{Funding}
Academy of Finland (298463, 318082 Flagship PREIN); Agence Nationale de la Recherche (ANR-15-IDEX- 0003, ANR-17-EURE-0002).	
\bibliographystyle{ieeetr}
\bibliography{Single_Pixel_Imaging}

\end{document}